\documentclass[aps,prd,preprint,eqsecnum,tightenlines,floats,
groupedaddress,showpacs]{revtex4}
\usepackage{amssymb} 
\usepackage{hyperref} 
\usepackage{graphicx}
\usepackage{subfigure}

\begin{document}

\preprint{FERMILAB-PUB-15-480-A-AE-PPD}

\date{November 17, 2015}

\title{Evolution of velocity dispersion along cold collisionless flows}

\author{Nilanjan Banik$^{a,b}$ and Pierre Sikivie$^a$}

\affiliation{$^a$ Department of Physics, University of Florida, 
Gainesville, FL 32611, USA\\
$^b$ Fermi National Accelerator Laboratory, Batavia, IL 60510, USA}

\begin{abstract}

The infall of cold dark matter onto a galaxy produces cold 
collisionless flows and caustics in its halo.  If a signal 
is found in the cavity detector of dark matter axions, the 
flows will be readily apparent as peaks in the energy spectrum 
of photons from axion conversion, allowing the densities, 
velocity vectors and velocity dispersions of the flows to 
be determined.  We discuss the evolution of velocity dispersion 
along cold collisionless flows in one and two dimensions.  A 
technique is presented for obtaining the leading behaviour of 
the velocity dispersion near caustics.  The results are used to 
derive an upper limit on the energy dispersion of the Big Flow 
from the sharpness of its nearby caustic, and a prediction for 
the dispersions in its velocity components.

\end{abstract}

\maketitle

\section{Introduction}

Among the outstanding problems in science today is the identity of the
dark matter of the universe \cite{PDM}.  The existence of dark matter is
implied by a large number of observations, including the dynamics of galaxy
clusters, the rotation curves of individual galaxies, the abundances of light
elements, gravitational lensing, and the anisotropies of the cosmic microwave
background radiation.  The energy fraction of the universe in dark matter is
observed to be 26\% \cite{Planck}.  The dark matter must be non-baryonic, cold
and collisionless.  {\it Non-baryonic} means that the dark matter is not made
of ordinary atoms and molecules.  {\it Cold} means that the primordial velocity
dispersion of the dark matter particles is sufficiently small, less than of order
$10^{-7}~c$ today, so that it may be set equal to zero as far as the formation
of large scale structure and galactic halos is concerned.  {\it Collisionless}
means that the dark matter particles have, in first approximation, only
gravitational interactions.  Particles with the required properties are
referred to as `cold dark matter' (CDM).  The leading CDM candidates are
weakly interacting massive particles (WIMPs) with mass in the 100 GeV range,
axions with mass in the $10^{-5}$ eV range, and sterile neutrinos with mass
in the keV range.

One approach to the identification of dark matter is to attempt to detect 
dark matter particles in the laboratory.  WIMP dark matter can be searched 
for on Earth by looking for the recoil of nuclei that have been struck by 
a WIMP \cite{Baudis}. Axion dark matter can be searched for by looking 
for the conversion of axions to photons in an electromagnetic cavity 
permeated by a strong magnetic field \cite{axdetth,axdetex}.  The 
spectrum of photons produced in the cavity is directly related to 
the axion energy spectrum in the laboratory since energy is conserved 
in the conversion process:             
\begin{equation}
h \nu = E_a = m_a c^2 + {1 \over 2} m_a v^2
\label{encon}
\end{equation}
where $\nu$ is the frequency of a photon produced by axion to photon
conversion, $E_a$ the energy of the axion that converted into it,
$m_a$ the axion mass and $\vec{v}$ the velocity of the axion in the
rest frame of the cavity.  Since the velocity dispersion of halo  
axions is of order $10^{-3} c$, the width of the axion signal is 
of order $10^{-6} \nu$.  If, for example, the axion signal occurs
at 1 GHz, its width is of order 1 kHz.  On the other hand, the 
resolution with which the signal can be spectrum analyzed is the 
inverse of the time over which it is observed \cite{hires}.  If 
the signal is observed for 100 seconds, for example, the achievable 
resolution is 0.01 Hz.  Thus, under the example given, the kinetic 
energy spectrum of halo axions is resolved into $10^5$ bins.  The 
Earth's rotation changes the velocities relative to the laboratory 
frame by amounts of order 1 m/s in 100 s, and therefore introduces 
D\"oppler shifts of order $\delta_D \nu \sim$ (300 km/s)(1 m/s) 
$\nu /c^2 \simeq 0.3 \cdot 10^{-11} \nu$, which is 0.003 Hz in the 
example given.  If the data taking runs are much longer than 100 s, 
the resolution is limited by the D\"oppler shifts.  However, for a 
given velocity vector in the rest frame of the Galaxy, the D\"oppler 
shifts may be removed and in that case the resolution can be much 
higher than 0.01 Hz.

Narrow peaks in the velocity spectrum of dark matter on Earth are expected
because a galactic halo grows continuously by accreting the dark matter
that surrounds it.  The infalling dark matter produces a set of flows
in the halo since the dark matter particles oscillate back and forth
many times in the galactic gravitational potential well before they are
thermalized by gravitational scattering off inhomogeneities in the galaxy
\cite{velpeak}.  The flows are cold and collisionless and therefore produce
caustics \cite{crdm,sing,inner,rob}.  Caustics are surfaces in physical 
space where the density is very high.  At the location of a caustic, a flow
``folds back" in phase space.  Each flow has a local density and velocity 
vector, and produces a peak with the corresponding properties in the energy 
spectrum of photons from axion conversion in a cavity detector.  All three 
components of a flow's velocity vector can be determined by observing the 
peak's frequency and D\"oppler's shift as a function of time of day and 
time of year \cite{modul}.  Thus an axion dark matter signal would be a 
rich source of information on the formation history of the Milky Way halo.  
Moreover, the flows are relevant to the search itself, before a signal is 
found, because a narrow prominent peak may have higher signal to noise than 
the full signal.

Motivated by these considerations, the self-similar infall model of galactic 
halo formation \cite{FGB} was used to predict the densities and velocity 
vectors of the Milky Way flows \cite{STW,MWhalo}.  The original model was 
generalized to include angular momentum for the infalling particles.  The 
flow  properties near Earth are sensitive to the dark matter angular momentum 
distribution because the angular momentum distribution determines the structure 
and location of the halo's inner caustics \cite{inner}. If the dark matter 
particles fall in with net overall rotation, the inner caustics are rings.  
The self-similar infall model predicts the radii of the caustic rings.  The 
caustic rings produce bumps in the galactic rotation curve at those radii.  
Mainly from the study of galactic rotation curves, but also from other data, 
evidence was found for caustic rings of dark matter at the locations predicted 
by the self-similar infall model.  The evidence is summarized in ref.\cite{MWhalo}.  
The model of the Milky Way halo that results from fitting the self-similar 
infall model to the data is called the ``Caustic Ring Model".  It is a 
detailed description of the full phase-space structure of the Milky Way 
halo \cite{MWhalo}.

The caustic ring model predicts that the local velocity distribution on
Earth is dominated by a single flow, dubbed the ``Big Flow" \cite{MWcr}.
The reason for this is our proximity to a cusp in the 5th caustic
ring in the Milky Way.  Up to a two-fold ambiguity, the Big Flow has 
a known velocity vector; see Section V.  Its density on Earth is 
estimated to be of order $1.5 \cdot 10^{-24}$ gr/cm$^3$, i.e. a factor 
three larger than typical estimates ($0.5 \cdot 10^{-24}$ gr/cm$^3$) 
found in the litterature for the total dark matter density on Earth.  
The existence of the Big Flow provides strong additional motivation
for high resolution analysis of the output of the cavity detector, 
since it produces a prominent narrow peak in the energy spectrum.  
It is desirable to have an estimate of the width of that peak since 
this determines the signal to noise ratio of a high resolution search 
for it.  The width of the peak is the energy dispersion of the 
Big Flow.  One of our main goals is to place an upper limit on the 
energy dispersion of the Big Flow from the observed sharpness of 
the 5th caustic ring.  More generally we want to study the evolution 
of velocity dispersion along cold collision flows, the relation between 
velocity dispersion and the distance scale over which caustics are 
smoothed, and the behavior of velocity dispersion very close to a 
caustic.  Our results may be relevant to other cold collisionless 
flows, in particular the streams of stars that result from the tidal 
disruption of galactic sattelites, such as the Sagittarius Stream 
\cite{Sagit}.

In Section II, we study the evolution of velocity dispersion along 
a cold collisionless flow in one dimension.  In Section III, we do 
the same for an axially symmetric flow in two dimensions.  In Section 
IV we present a technique for obtaining the leading behavior of velocity 
dispersion near caustics.  We apply it to fold caustics and cusp caustics.
In Section V, we use our results to derive an upper limit on the energy 
disperion of the Big Flow from the sharpness of the 5th caustic ring and 
make a prediction for the relative dispersions of its velocity components. 
Section VI provides a brief summary.

\section{A cold flow in one dimension}

In this section, we study how velocity dispersion changes along a cold 
collisionless flow in one dimension.  We consider a large number of 
particles moving in a time-independent potential $V(x)$ and forming 
a stationary flow.  We first discuss the case where the velocity 
dispersion vanishes and then the case where the velocity dispersion
is small.

\subsection{Zero velocity dispersion}

In the case of zero velocity dispersion, all particles have 
the same energy
\begin{equation}
E = {m \over 2} v^2 + V(x)~~\ .
\label{energy}
\end{equation}
Hence their velocity at location $x$ is 
\begin{equation}
v(x) = \pm \sqrt{{2 \over m}(E - V(x))}~~\ .
\label{velx}
\end{equation}
For the sake of definiteness, we assume that the particles are 
bound to a minimum of $V(x)$, going back and forth between $x_1$ 
and $x_2$, defined by $E = V(x_1) = V(x_2)$.  For $x_1 < x < x_2$ 
there are two flows, one with $v(x) > 0$ and one with $v(x) < 0$. 
There are no flows at $x < x_1$ or $x > x_2$.  Since the overall 
flow is stationary (${\partial n \over \partial t} = 0$), the 
continuity equation
\begin{equation}
{\partial n \over \partial t} + 
{\partial \over \partial x} (nv) = 0~~\ ,
\label{coneq}
\end{equation}
implies that the densities of both the left- and right-moving 
flows equal
\begin{equation}
n(x) = {J \over \sqrt{{2 \over m}(E - V(x))}}~~\ ,
\label{denx}
\end{equation}
where $J$ is a constant. $J$ is the flux (number of particles per
unit time) in the left- and the right-moving flows.  

There are two caustics, one at $x_1$ and the other at $x_2$.  The 
caustics are simple fold ($A_2$) catastrophes.  For $x$ near $x_1$ 
\begin{equation}
V(x) = E + (x - x_1) {dV \over dx}(x_1) + {\cal O}(x - x_1)^2
\label{Taylor}
\end{equation}
with ${dV \over dx}(x_1) < 0$.  Hence 
\begin{equation}
n(x) = {J \over \sqrt{- {2 \over m} {dV \over dx}(x_1)}}{1 \over \sqrt{x - x_1}}
\label{caus1}
\end{equation}
as $x$ approaches $x_1$ from above.  Likewise
\begin{equation}
n(x) = {J \over \sqrt{+ {2 \over m}{dV \over dx}(x_2)}}{1 \over \sqrt{x_2 - x}}
\label{caus2}
\end{equation}
when $x$ approaches $x_2$ from below.

\subsection{Small velocity dispersion}

Next we consider the same flow but with a small energy dispersion 
$\delta E$.  We assume that the energy distribution ${dN \over dE}(E)$ 
is narrowly peaked about its average $\bar{E}$.  $\delta E$ is defined 
as usual by 
\begin{equation}
\delta E = <(E - \bar{E})^2>^{1 \over 2}~~\ .
\label{delE}
\end{equation}
Brackets indicate averaging over the energy distribution. 
At location $x$, the two flows have average velocity
\begin{equation}
\bar{v}(x) \simeq v(x,\bar{E}) = 
\pm \sqrt{{1 \over 2m}(\bar{E} - V(x))}
\label{barv}
\end{equation}
and velocity dispersion
\begin{equation}
\delta v(x) = <(v(x,E) - \bar{v}(x))^2>^{1 \over 2} \simeq 
<\left((E - \bar{E}) {\partial v \over \partial E}(x,\bar{E}) \right)^2>^{1 \over 2}
= {\delta E \over m |v(x,\bar{E})|}~~~\ .
\label{delv}
\end{equation}
We assumed that ${dN \over dE}(E)$ goes to zero rapidly as 
$|E - \bar{E}|$ increases, as is the case e.g. for a Gaussian 
distribution.  Eq.~(\ref{barv}) is exact then in the limit 
$\delta E \rightarrow 0$.  Also Eq.~(\ref{delv}) is exact in 
that limit provided that, in addition, 
${\partial E \over \partial v}|_x = m v(x, \bar{E})$ does not 
vanish, i.e. that one is not close to a caustic.

The density of each of the flows is given by Eq.~(\ref{denx}),
with $E$ replaced by $\bar{E}$.  The phase-space density of
the flow is therefore
\begin{equation}
{\cal N} = {n(x) \over m \delta v(x)} \simeq {J \over \delta E}~~ \ .
\label{Liou}
\end{equation}
It is independent of $x$, as required by Liouville's theorem.

Velocity dispersion smoothes the caustics.  Indeed, 
when $\delta E \neq 0$, the caustics are spread over a 
thickness ($j = 1,2$)
\begin{equation}
\delta x_j = {\delta E \over \Big|{dV \over dx}(x_j)\Big|}~~\ .
\label{thick}
\end{equation}
The density reaches at the caustics a maximum value
\begin{equation}
n_{{\rm max},j} \sim 
\sqrt{m \over 2 \Big|{dV \over dx}(x_j)\Big|}
{J \over \sqrt{\delta x_j}} = \sqrt{m \over 2 \delta E} J~~\ .
\label{maxden}
\end{equation}
Also the velocity dispersion reaches a maximum value
\begin{equation}
\delta v_{{\rm max},j} \sim \sqrt{\delta E \over 2m}~~\ .
\label{maxveldis}
\end{equation}
It is interesting that $n_{\rm max}$ and $\delta v_{\rm max}$ are
the same at one caustic as at the other.  Eq.~(\ref{maxveldis}) 
follows from Eq.~(\ref{maxden}) and Liouville's theorem.  It also 
follows from the fact that $\bar{v} = 0$ at the caustics, 
so that $\delta E \sim {m \over 2} (\delta v_{\rm max})^2$
there.

\section{An axisymmetric cold flow in two dimensions}

In this section we study a stationary, axisymmetric, cold, 
collisioness flow of particles moving in two dimensions in an 
axisymmetric time-independent potential $V(r)$.  Again, we 
discuss first the flow with vanishing velocity dispersion, and 
then the flow with a small velocity dispersion.  

\subsection{Zero velocity dispersion}

Consider a flow of particles moving in a plane, under the influence
of a potential $V(r)$. $(r,\phi)$ are polar coordinates in the plane.
All particles have the same energy $E$ and the same angular momentum 
$L$.  Hence the velocity field
\begin{equation}
\vec{v}(r,\phi) = v_r(r)\hat{r} + v_\phi(r) \hat{\phi}
\label{2vel}
\end{equation}
with 
\begin{equation}
v_\phi(r) = {L \over mr} ~~~{\rm and}~~~
v_r(r) = \pm \sqrt{{2 \over m}(E - V_{\rm eff}(r))}
\label{velcom}
\end{equation}
where 
\begin{equation}
V_{\rm eff} = V(r) + {L^2 \over 2 m r^2}~~~\ .
\label{effpot}
\end{equation}
Provided $L \neq 0$ the particles have a non-zero distance of 
closest approach $a$: $E = V_{\rm eff}(a)$.  Let us assume they 
also have a turnaround radius $R$, with $E = V_{\rm eff}(R)$ and 
$R>a$.  There are two flows for $a < r < R$.  We call them 
the "in" ($v_r < 0$) and "out" ($v_r > 0$) flows.   There are  
no flows for $r < a$ or $r > R$.  

Since the flow is stationary and axisymmetric, the continuity 
equation implies that the density of particles of both 
the in and out flows is
\begin{equation}
n(r) = {J \over r |v_r(r)|}~~~
\label{2den}
\end{equation}
where $J$ is a constant.  $J$ is the number of particles per unit 
time and per radian.  There are simple fold caustics at $r = a$ and 
$r = R$.  When $r$ approaches $a$ from above
\begin{equation}
n(r) = {J \over a \sqrt{- {2 \over m} {d V_{\rm eff} \over dr}(a)}} 
{1 \over \sqrt{r-a}}~~\ ,
\label{ca}
\end{equation}
whereas
\begin{equation}
n(r) = {J \over R \sqrt{+ {2 \over m} {d V_{\rm eff} \over dr}(R)}} 
{1 \over \sqrt{R-r}}
\label{cR}
\end{equation}
when $r$ approaches $R$ from below.

\subsection{Small velocity dispersion}

We now consider the same flow as in the previous subsection but with a 
Gaussian distribution of energy and angular momentum of the form
\begin{equation}
{d^2 N \over dE dL} = {N \over 2 \pi \sigma_E \sigma_L}
e^{-{1 \over 2}({E - \bar{E} \over \sigma_E})^2 
-{1 \over 2}({L - \bar{L} \over \sigma_L})^2}~~\ .
\label{Gauss}
\end{equation}
For this distribution $\delta E = \sigma_E$, $\delta L = \sigma_L$, and 
\begin{equation}
<(E - \bar{E})(L - \bar{L})>~=~0~~\ .
\label{diag}
\end{equation}
The most general Gaussian would allow $ <(E - \bar{E})(L - \bar{L})> \neq 0$.  
However our main interest is the evolution of the velocity dispersion of 
flows of cold dark matter particles falling onto a galactic halo and sloshing 
back and forth thereafter.  We now argue that Eq.~(\ref{diag}) is a good
approximation for that case. 

The primordial velocity dispersion $\delta v_p$ of a flow of cold dark matter 
particles is negligibly small.  By primordial velocity dispersion, we mean the 
velocity dispersion that the particles have in the absence of structure formation.  
$\delta v_p \sim 10^{-17}$ is typical of axions, $\delta v_p \sim 10^{-12}$ for 
WIMPs, and $\delta v_p \sim 10^{-8}$ for sterile neutrinos.  The main contributions 
to the velocity dispersion of a flow of cold dark matter particles falling onto a 
galaxy are instead from gravitational scattering off inhomogeneities in the galaxy 
(such as globular clusters and molecular clouds) and from the growth by gravitational 
instability of small scale density perturbations in the flow itself.  When a process 
produces a velocity dispersion $\delta v$, the associated energy dispersion is of 
order $\delta E \sim m v \delta v$ where $v$ is the velocity of the flow in the 
galactic reference frame when the process occurs, and the associated angular 
momentum dispersion is of order $\delta L \sim m D \delta v$ where $D$ is the 
distance from the galactic center where the process occurs.  So, $\delta E$ 
and $\delta L$ are not independent quantities but related by 
\begin{equation}
\delta L = \delta E~{D_{\rm av} \over v_{\rm av}}
\label{relat}
\end{equation}
where $v_{\rm av}$ is an average flow velocity, say 300 km/s for our
Milky Way galaxy, and $D_{\rm av}$ is an average distance from the 
galactic center where the flow acquired velocity dispersion.  We may 
only give a rough guess for the order of magnitude of $D_{\rm av}$, 
perhaps 100 kpc for our galaxy.  In view of Eq.~(\ref{relat}) we define an 
overall flow velocity dispersion $\sigma_v$:  $\delta L = m D_{\rm av} \sigma_v$
and $\delta E = m v_{\rm av} \sigma_v$.

Furthermore, in the limit where the galaxy has no angular momentum 
($\bar{L} = 0$), there is no preference for the many events that produce 
velocity dispersion to increase or decrease $<(E - \bar{E})(L - \bar{L})>$ 
since this quantity is odd under $L \rightarrow - L$.  Therefore 
$<(E - \bar{E})(L - \bar{L})>$ is proportional to $\bar{L}$.  Disk 
galaxies have angular momentum but, relative to their size and typical 
velocities, that angular momentum is small.  Indeed all dimensionless 
measures of galactic angular momentum have values much less than one.  
One such measure is the galactic spin parameter \cite{Peeb}
\begin{equation} \lambda = {{\cal L} 
|{\cal E}|^{1 \over 2} \over G {\cal M}^{5 \over 2}} 
\label{lam} 
\end{equation} 
where ${\cal L}$ is the angular momentum of the galaxy, ${\cal E}$ 
its net mechanical (kinetic plus gravitational) energy and ${\cal M}$ 
its mass. $G$ is Newton's gravitational constant.  A typical value is 
$\lambda \sim 0.05$.  Another dimensionless measure of galactic angular 
momentum is the ratio $a/R$ of caustic ring radius $a$ to turnaround 
radius $R$ for the flows of dark matter particles in the halo.  A typical 
value is $a/R \sim 0.1$ \cite{MWhalo}.  Finally, the dimensionless number 
that controls the amount of galactic angular momentum in the Caustic Ring 
Model is $j_{\rm max}$.  A typical value is $j_{\rm max} \sim$ 0.2.  Since 
$<(E - \bar{E})(L - \bar{L})>$ is proportional to galactic angular 
momentum, and galactic angular momentum is of order 0.1 in dimensionless 
units, $<(E - \bar{E})(L - \bar{L})>$ is suppressed relative to 
$\delta E ~ \delta L$ by a similar factor of order 0.1.  To 
simplify our calculations, we set $<(E - \bar{E})(L - \bar{L})>$ 
= 0 as a first approximation.

Provided $\delta E$ and $\delta L$ are sufficiently small, we may 
within the support of the ${d^2 N \over dE dL}$ distribution express
small deviations $dE = E - \bar{E}$ and $dL = L - \bar{L}$ of the 
particle energy and angular momentum from its average values as 
linear functions of small deviations $dv_r = v_r - \bar{v}_r$ and 
$dv_\phi = v_\phi - \bar{v}_\phi$ of the velocity components from their 
average values at a given position $(r,\phi)$.  Henceforth we set $m = 1$
to avoid cluttering the equations unnecessarily.  Eqs.~(\ref{velcom}) imply
\begin{equation}
dE = v_r dv_r + v_\phi dv_\phi ~~~{\rm and}~~~ dL = r dv_\phi~~~\ .
\label{1stvar}
\end{equation}
Therefore the exponent in Eq.~(\ref{Gauss}) may be rewritten using
\begin{equation}
(dE~~dL)
\left( \begin{array} {cc} {1 \over \sigma_E^2} ~~ & ~~0 \\
0~~ & ~~ {1 \over \sigma_L^2} \end{array} \right)
\left( \begin{array} {c} dE \\ dL \end{array} \right) = 
(dv_r ~~ dv_\phi) 
\left( \begin{array} {cc} 
{v_r^2 \over \sigma_E^2} ~~ & ~~{v_r v_\phi \over \sigma_E^2} \\
{v_r v_\phi \over \sigma_E^2} ~~ & ~~ {v_\phi^2 \over \sigma_E^2} 
+ {r^2 \over \sigma_L^2} \end{array} \right) 
\left( \begin{array} {c} dv_r \\ dv_\phi \end{array} \right)~~\ .
\label{matrix}
\end{equation}
We rotate
\begin{equation}
\left( \begin{array} {c} dv_r \\ dv_\phi \end{array} \right) = 
\left( \begin{array} {cc} \cos \theta ~~ & ~~ \sin \theta \\
- \sin \theta ~~ & ~~ \cos \theta \end{array} \right) 
\left( \begin{array} {c} dv_1 \\ dv_2 \end{array} \right) 
\label{rotate}
\end{equation}
so as to diagonalize the 2x2 matrix on the RHS of Eq.~(\ref{matrix}).  
Provided
\begin{equation}
\tan 2 \theta = {2 v_r v_\phi \over 
v_\phi^2 + \left( {\sigma_E \over \sigma_L} r \right)^2 - v_r^2}
\label{theta}
\end{equation}
we have 
\begin{equation} 
\left({dE \over \sigma_E}\right)^2 + \left({dL \over \sigma_L}\right)^2
= \left({dv_1 \over \sigma_1}\right)^2 + \left({dv_2 \over \sigma_2}\right)^2
\label{quadform}
\end{equation}
where 
\begin{equation} 
{1 \over (\sigma_{1 \atop 2})^2} = 
{1 \over 2} \left({v_r^2 + v_\phi^2 \over \sigma_E^2}
+ {r^2 \over \sigma_L^2} \right) \mp
\sqrt{{1 \over 4} \left({v_r^2 + v_\phi^2 \over \sigma_E^2}
+ {r^2 \over \sigma_L^2} \right)^2 - {r^2 v_r^2 \over \sigma_E^2 \sigma_L^2}}~~\ .
\label{veldis2}
\end{equation}
This implies 
\begin{equation}
\sigma_1 ~ \sigma_2 = {\sigma_E ~ \sigma_L \over r |v_r(r)|}~~\ .
\label{relation}
\end{equation}
At a given location, the velocity distribution is 
\begin{equation}
{d^2 N \over dv_r dv_\phi} = {d^2 N \over dE dL} 
|\det\left({\partial(E,L) \over \partial(v_r, v_\phi)}\right)|
= r |v_r(r)| {d^2 N \over dE dL}~~\ .
\label{dist2}
\end{equation}
We have
\begin{eqnarray}
(\delta v_r)^2 &=& 
\cos^2\theta (\sigma_1)^2 + \sin^2\theta (\sigma_2)^2\nonumber\\
(\delta v_\phi)^2 &=& \sin^2\theta (\sigma_1)^2 
+ \cos^2\theta (\sigma_2)^2\nonumber\\
<dv_r~dv_\phi> &=& (\sigma_2^2 - \sigma_1^2) \sin\theta~\cos\theta~~\ .
\end{eqnarray}
Liouville's theorem is satisfied since the phase space density
\begin{equation}
{\cal N} = {n(r) \over \sigma_1 \sigma_2} = 
{n(r) r |v_r(r)| \over \sigma_E \sigma_L} = {J \over \sigma_E \sigma_L}
\label{Liou2}
\end{equation}
does not depend on $r$.  

At the caustics, $\theta \rightarrow 0$ since $v_r \rightarrow 0$.
$\sigma_2$ becomes $\delta v_\phi$ and remains finite:
\begin{equation}
\sigma_2 \rightarrow {1 \over \sqrt{\left({L \over r_c \sigma_E}\right)^2 
+ \left({r_c \over \sigma_L}\right)^2}} \equiv \delta v_\phi(r_c)
\label{limdv2}
\end{equation}
with $r_c = a$ or $R$.  $\sigma_1$ becomes $\delta v_r$ and is large.  
According to Eqs.~(\ref{veldis2}) and (\ref{2den}), $\delta v_r$ 
and $n(r)$ become infinite.  However, those equations cannot be 
used right at the caustics since they neglect second order terms 
in Eqs.~(\ref{1stvar}), and this is inaccurate when $v_r \rightarrow$ 0.

We may use Eq.~(\ref{Liou2}) to estimate the maximum density and 
velocity dispersion at the caustics since that equation follows 
directly from Liouville's theorem.  The divergence at the caustics 
is cut off by the fact that $v_r \sim \delta v_r$ there.  Through 
Eqs.~(\ref{relation}) and (\ref{2den}), this implies
\begin{equation}
\delta v_{r,{\rm max}}|_{r_c}  
\sim \sqrt{\sigma_E \sigma_L \over r_c  \delta v_\phi(r_c)}
\label{maxveldis2}
\end{equation}
and 
\begin{equation}
n_{\rm max}|_{r_c} \sim J 
\sqrt{\delta v_\phi(r_c) \over r_c \sigma_E \sigma_L}~~\ .
\label{nmax}
\end{equation}
At the inner caustic
\begin{equation}
\delta v_\phi(a) = 
{1 \over \sqrt{\left({L \over a \sigma_E}\right)^2 +
\left({a \over \sigma_L}\right)^2}} = 
{1 \over \sqrt{\left({v_\phi(a) \over v_{\rm av} \sigma_v}\right)^2
+ \left({a \over D_{\rm av} \sigma_v}\right)^2}}\simeq 
\sigma_v {v_{\rm av} \over v_\phi(a)}
\label{vpa}
\end{equation}
in the limit of small angular momentum ($a << R$) since 
$v_\phi(a) = {L \over a}$ is of order $v_{\rm av}$ whereas
$D_{\rm av}$ is of order $R$ and therefore much larger than 
$a$.  Hence
\begin{equation}
\delta v_{r,{\rm max}}|_a \sim 
\sqrt{{1 \over a} D_{\rm av} v_\phi(a) \sigma_v} = 
{1 \over a} \sqrt{L \sigma_L}
\label{dvmaxa}
\end{equation}
and 
\begin{equation}
n_{\rm max}|_a \sim {J \over \sqrt{L \sigma_L}}~~~\ .
\label{nmaxa}
\end{equation}
Eqs.~(\ref{dvmaxa}) and (\ref{nmaxa}) show that the properties
of the inner caustic are controlled by the angular momentum 
distribution in the small angular momentum limit.

In contrast, at the outer caustic, 
\begin{equation}
\delta v_\phi(R) = 
{1 \over \sqrt{\left({L \over R \sigma_E}\right)^2 + 
\left({R \over \sigma_L}\right)^2}} = 
{1 \over \sqrt{\left({a v_\phi(a) \over R v_{\rm av} \sigma_v}\right)^2
+ \left({R \over D_{\rm av} \sigma_v}\right)^2}} \simeq
\sigma_v {D_{\rm av} \over R}
\label{vpR}
\end{equation}
since $a << R$ in the limit of small angular momentum.  Hence
\begin{equation}
\delta v_{r,{\rm max}}|_R \sim \sqrt{v_{\rm av} \sigma_v} 
= \sqrt{\delta E}
\label{dvmaxR}
\end{equation}
and 
\begin{equation}
n_{\rm max} \sim {J \over R \sqrt{\delta E}}~~\ .
\label{nmaxR}
\end{equation}
The properties of the outer caustic are controlled by the 
energy distribution in the small angular momentum limit.

\section{Velocity dispersion near a caustic}

The local velocity dispersion becomes large when a caustic is 
approached since the physical space density becomes large but 
the phase-space density remains constant.  In this section we 
present a technique for deriving the leading behavior of the 
velocity dispersion near caustics.  We apply it to the case of 
fold caustics and of cusp caustics.

\subsection{Velocity dispersion near a fold caustic}

The fold caustic is described by the simplest ($A_2$) 
of catastrophes.  It is the only caustic possible in 
one dimension.  In one dimension it occurs at a point. 
In two dimensions it occurs on a line, and in three 
dimensions it occurs on a surface, with the dimensions 
parallel to the line or surface playing spectator roles 
only.

Consider a flow of particles moving in one dimension in the 
potential $V(x) = g x$ where $g$ is a constant acceleration.  
We set the mass of the particles equal to one, as in the 
previous section.  We assume the flow to be stationary for 
simplicity.  This is not an essential assumption. In the 
limit of zero velocity dispersion, the flow at a given time 
is described by the map
\begin{equation}
x(\tau) = x_0 - {1 \over 2} g \tau^2
\label{fmap}
\end{equation}
which gives the position of the particles as a function of their 
age.  We define the age $\tau$ of a particle as minus the time at 
which it passes through its maximum $x$ value.  The velocity is 
\begin{equation}
v(\tau) = {\partial x \over \partial \tau} = - g \tau~~\ .
\label{fvel}
\end{equation}
At position $y$ there are two flows if $y < x_0$, and no 
flow if $y > x_0$.  For $y < x_0$, the particles at position 
$y$ have ages $\tau = \pm \sqrt{{2 \over g}(x_0 - y)}$.
The density of each of the two flows is
\begin{equation}
n(y) \equiv {d N \over d y} (y) = 
{dN \over d\tau} |{d \tau \over dy}| 
= {dN \over d\tau} {1 \over \sqrt{2 g (x_0 - y)}}~~\ .
\label{fden}
\end{equation}
$N$ is number of particles, as before.  The fold caustic 
is located at $y = x_0$.

We now allow the particles to have a distribution 
${d N \over d x_0}$ of $x_0$ values, with average 
$\bar{x}_0$ and a small dispersion $\delta x_0$.  
Since the particles at a given location $y$ have a 
distribution of $x_0$ values, they also have a 
distribution of ages. Small deviations $d x_0$ 
and $d \tau$ from the average $x_0$ and the average 
age at a given location are related by 
\begin{equation}
0 = dx = d x_0 - g \tau d\tau~~\ .
\label{frel}
\end{equation}
Hence the dispersion in ages at a given location is 
\begin{equation}
\delta \tau = \sqrt{<(d \tau)^2>} =  
\sqrt{<\left({d x_0 \over g \tau}\right)^2>}
= {1 \over g |\tau|} \delta x_0~~\ .
\label{cdt}
\end{equation}
Brackets indicate averages over the distribution 
${dN \over d x_0}$. Small deviations in velocity 
are related to small deviations in age by  $dv = - g d\tau$.  
Hence the velocity dispersion at position $y$ is
\begin{equation}
\delta v(y)  = g \delta \tau =  
{\delta x_0 \over |\tau|} = 
\delta x_0 \sqrt{g \over 2 (x_0 - y)}~~\ .
\label{fdvel}
\end{equation}
The phase space density ${\cal N} = n/\delta v$ has no 
singularity at the caustic, in accordance with Liouville's 
theorem.  Although everything we found here was already 
obtained in Section II, the present approach is more 
efficient is obtaining the characteristic behaviour of 
the velocity dispersion right at the caustic.  The 
technique can be readily applied to more complicated 
cases, such as the cusp caustic which we discuss next.

\subsection{Velocity dispersion near a cusp caustic}

The cusp caustic is described by the $A_3$ catastrophe, 
the next simplest after the fold catastrophe. It can only 
exist in two dimensions or higher.  In two dimensions, it 
occurs at a point.  In three dimensions, it occurs along a 
line with the dimension parallel to the line playing a 
spectator role only.  

A cusp caustic appears in the map
\begin{eqnarray}
z &=& z_0 + b \alpha \tau\nonumber\\
x &=& x_0 - c \tau - {1 \over 2}s \alpha^2
\label{cmap}
\end{eqnarray}
giving the positions $(x,z)$ of particles in a flow from right 
to left with the particles on the right and top moving downwards 
and the particles on the right and bottom moving upwards.  See 
Fig.~1.  For simplicity, we take the flow to be stationary, and 
the acceleration to vanish everywhere.  The particles are labeled 
by $(\alpha, \tau)$ where $\tau$ is an age parameter, defined as 
minus the time a particle crosses the $z=0$ axis. $\alpha$ labels 
the different trajectories.  The velocities are 
\begin{eqnarray}
v_z &=& {\partial z \over \partial \tau} = b \alpha\nonumber\\
v_x &=& {\partial x \over \partial \tau} = - c~~\ .
\label{cvel}
\end{eqnarray}
The density of the flow is 
\begin{equation}
d(x,z) \equiv {d^2 N \over dx dz} = 
{d^2 N \over d\alpha d\tau} {1 \over |D(\alpha,\tau)|}
\label{cden}
\end{equation}
where 
\begin{equation}
D(\alpha,\tau) = 
det \left({\partial(x,z) \over \partial (\alpha,\tau)}\right)
= b ( - s \alpha^2 + c \tau)~~\ .
\label{jac}
\end{equation}
Caustics occur where $D(\alpha,\tau) = 0$, i.e. where the map is 
singular.  The equation $D(\alpha,\tau) = 0$  defines the curve
\begin{equation}
x = x_0 - {3 \over 2} \left({s c^2 \over b^2}\right)^{1 \over 3}
|z - z_0|^{2 \over 3}~~\ ,
\label{ccur}
\end{equation}
shown by the solid line in Fig.~1.  There are three flows 
at every location to the left of the curve whereas there is 
only one flow in the region to its right.  The number of flows 
at a location $(x,z)$ is the number of $(\alpha,\tau)$ values that 
solve Eqs.~(\ref{cmap}).  The curve is the location of a fold caustic.  
The cusp caustic is the special point $(x_0,z_0)$.  When the curve 
is traversed starting on the side with three flows, two of the flows 
disappear.  At any point other than $(x_0,z_0)$, the density and the 
velocity dispersion of those two flows behave as described in the 
previous subsection, i.e. the density diverges as 
${1 \over \sqrt{h}}$ where $h$ is the distance to the curve and 
the dispersion in the velocity component perpendicular to the curve 
also diverges as ${1 \over \sqrt{h}}$.  The dispersion in the velocity 
component parallel to the curve remains finite.  

In this subsection, we are interested in the behavior of the density 
and velocity dispersion at the cusp.  A complete answer can be given 
by obtaining the inverse of the map in Eqs. (\ref{cmap}).  This involves 
solving a third order polynomial equation, and inserting the resulting 
functions $\alpha(x,z)$ and $\tau(x,z)$ into the RHS of Eqs.~(\ref{jac})
and (\ref{cden}).  Here we content ourselves with the behavior of the 
density and the velocity dispersion ellipse when the cusp is approached 
from particular directions.  If the cusp is approached along the $z=z_0$ 
axis from the right ($x>x_0$), the single flow there has $\alpha = 0$
and $\tau = - (x - x_0)/c$.  The density of that flow is
\begin{equation}
d(x,z_0) = {d^2 N \over d\alpha d\tau} {1 \over b |x - x_0|}~~\ .
\label{frr}
\end{equation}
If the cusp is approached along the $x = x_0$ axis, from the top or from 
the bottom, the single flow there has $\tau = - {s \over 2 c} \alpha^2$
and $z - z_0 = - {bs \over 2c} \alpha^3$.  The density is
\begin{equation}
d(x_0,z) = {d^2 N \over d\alpha d\tau}~{1 \over 3}
\left({2 \over b s c^2}\right)^{1 \over 3}
{1 \over |z - z_0|^{2 \over 3}}~~\ .
\label{frtb}
\end{equation} 
If the cusp is approached along the $z=z_0$ axis from the left ($x<x_0$)
there are three flows: i) $\alpha = 0$ and $\tau = (x_0 - x)/c$, 
ii) $\tau = 0$ and $\alpha = \sqrt{{2 \over s}(x_0 - x)}$, and 
iii) $\tau = 0$ and $\alpha = - \sqrt{{2 \over s}(x_0 - x)}$.
Flow i) has the same density as given in Eq.~(\ref{frr}) whereas 
flows ii) and iii) each have half the density given in Eq.~(\ref{frr}).

To obtain the velocity dispersions we consider a set of maps, as in 
Eq.~(\ref{cmap}), but with a distribution ${d^5 N \over dz_0~db~dx_0~dc~ds}$ 
of the constants $z_0$, $b$, $x_0$, $c$ and $s$ that appear there.  We 
assume that the distribution is narrowly peaked around average values 
of these constants.  We consider small deviations $dz_0$ ... $ds$ of
the constants from their average values plus small deviations $d\alpha$ 
and $d\tau$ in the flow parameters such that 
\begin{eqnarray}
dz &=& dz_0 + db \alpha \tau 
+ b d\alpha \tau + b \alpha d\tau = 0\nonumber\\
dx &=& dx_0 -dc \tau - c d\tau 
- {1 \over 2} ds \alpha^2 - s \alpha d\alpha = 0~~\ .
\label{crel}
\end{eqnarray}
Eqs.~(\ref{crel}) imply that, at a given physical point, 
the deviations in the flow parameters are given in terms 
of the deviations in the constants by 
\begin{equation}
\left( \begin{array} {cc}
b \tau ~~ & ~~ b \alpha \\
s \alpha ~~ & ~~ c \end{array} \right)
\left( \begin{array} {c} d \alpha\\ d \tau \end{array} \right)
= \left( \begin{array} {c} 
- dz_0 - \alpha \tau db \\ 
dx_0 - \tau dc - {1 \over 2} \alpha^2 ds \end{array} \right)~~\ .
\label{crel2}
\end{equation}
When approaching the cusp, $\alpha \rightarrow 0$ and $\tau \rightarrow 0$,
the RHS of Eq.~(\ref{crel2}) goes to $\left({- dz_0 \atop dx_0}\right)$, 
and therefore
\begin{equation}
\left( \begin{array} {c} d \alpha\\ d \tau \end{array} \right) = 
{1 \over b (c \tau - s \alpha^2)}
\left( \begin{array} {cc}
c~~ & ~~ - b \alpha \\
- s \alpha ~~ & ~~ b \tau \end{array} \right)
\left( \begin{array} {c} - d z_0 \\ d x_0\end{array} \right)~~\ .
\label{crel3}
\end{equation}
Likewise, when approaching the cusp, the deviations in the 
velocity components are
\begin{eqnarray}
dv_z &=&  b d\alpha = - {c dz_0 \over c \tau - s \alpha^2}\nonumber\\
dv_x &=& - dc~~\ .
\label{cveld}
\end{eqnarray}
Hence
\begin{equation}
\delta v_z = \sqrt{<(d v_z)^2>} = 
{c \delta z_0\over |c \tau - s \alpha^2|} ~~~,~~~
\delta v_x = \delta c~~\ .
\label{cveldis}
\end{equation}
If $d z_0$ and $dc$ are correlated 
\begin{equation}
<dv_z~dv_x> = 
{c \over c \tau - s \alpha^2} <dz_0~dc>~~\ .
\label{ccor}
\end{equation} 
For each of the flows at the cusp, the dispersion in the velocity 
component parallel to the axis ($\hat{x}$) of the cusp remains finite
whereas the dispersion in the component of velocity in the direction 
perpendicular ($\hat{z}$) to the axis of the cusp diverges.  This 
might have been expected since the flow folds in the direction 
perpendicular to the axis of the cusp.  The phase space densiy 
remains finite since the divergence of the physical density is 
canceled by the divergence of the velocity dispersion.

\section{Applications to the Big Flow}

In the Caustic Ring Model of the Milky Way halo, the dark 
matter density on Earth is dominated by a single cold flow, 
dubbed the `Big Flow', because of our proximity to a cusp 
in the 5th caustic ring in our galaxy.  In the Caustic Ring 
Model, there are two flows on Earth associated with the 5th 
caustic ring.  Their velocity vectors are \cite{MWcr} \cite{MWhalo}
\begin{equation}
\vec{v}_{5\pm} \simeq (505~\hat{\phi} \pm 120~\hat{r})~{\rm km/s}
\label{BFvel}
\end{equation}
where $\hat{\phi}$ is the unit vector in the direction of Galactic
rotation and $\hat{r}$ the unit vector in the radially outward
direction.  The density of the Big Flow on Earth is estimated to
be $1.5 \cdot 10^{-24}$ gr/cm$^3$. The density of the other flow 
associated with the 5th caustic ring, hereafter called the `Little Flow',  
is estimated to be $0.15 \cdot 10^{-24}$ gr/cm$^3$.  It is not known 
whether the Big Flow has velocity $\vec{v}_{5-}$ and the Little 
Flow has velocity $\vec{v}_{5+}$, or vice-versa.  The existence 
of the Big Flow provides strong additional motivation for high 
resolution analysis of the output of a cavity detector of dark 
matter axions, since it produces a prominent narrow peak in the
energy spectrum.  The width of the peak determines the signal 
to noise ratio of a high resolution search for it.  The width 
of the peak is the energy dispersion of the Big Flow.  Here we 
place an upper limit on the energy dispersion of the Big Flow 
from the observed sharpness of the 5th caustic ring.  The same 
upper limit applies to the Little Flow.

The best lower limit on the sharpness of the 5th caustic ring is obtained
by considering a triangular feature in the Infrared Astronomical Sattelite
(IRAS) map of the Galactic plane \cite{MWcr}.  The feature is in a direction 
tangent to the 5th caustic ring.  The gravitational fields of caustic rings 
of dark matter leave imprints upon the spatial distribution of ordinary
matter.  Looking tangentially to a caustic ring, from a vantage point in
the plane of the ring, one may have the good fortune of recognizing the
tricusp shape \cite{sing} of the cross-section of a caustic ring.  The 
IRAS map of the Galactic plane in the direction tangent to the 5th 
caustic ring shows such a feature. The relevant IRAS maps are posted 
at http://www.phys.ufl.edu/~sikivie/triangle.  The triangular feature 
is correctly oriented with respect to the galactic plane and the galactic 
center.  Its position is consistent within measurement errors with the 
position of the sharp rise in the Galactic rotation curve due to the 
5th caustic ring.  If the velocity dispersion of the flow forming the 
5th caustic ring were large, the triangular feature in the IRAS map 
would be blurred.  The sharpness of the triangular feature in the IRAS 
map implies that the 5th caustic ring is spread over a distance less 
than approximately 10 pc.  

\subsection{Upper limit on the Big Flow energy dispersion}

The particles forming a caustic ring fall in and out of the galaxy near 
the galactic plane.   Let $E$ and $L$ be respectively the energy and 
angular momentum of the particles that form the 5th caustic ring and 
are in the galactic plane.  We have 
\begin{equation}
E = {1 \over 2} v_r(r)^2 + {L^2 \over 2 r^2} + V(r)
\label{encona}
\end{equation}
where $v_r(r)$ is the radial velocity of the particles at 
galactocentric radius $r$ and $V(r)$  is the gravitational 
potential in which they move.  We will use 
\begin{equation}
V(r) = v_{\rm rot}^2 \ln(r)
\label{gravpot}
\end{equation}
consistent with a flat galactic rotation curve, 
with rotation velocity $v_{\rm rot}$.  For the Milky Way, 
$v_{\rm rot} \simeq$ 220 km/s.  The inner radius $a$ of a caustic 
ring is the distance of closest approach to the galactic center 
of the particles in the galactic plane.  Therefore
\begin{equation}
E = {L^2 \over 2 a^2} + V(a)~~\ .
\label{eqa}
\end{equation}
Small deviations $dE$, $dL$ and $da$ from the average 
values of $E$, $L$ and $a$ are related by 
\begin{equation}
dE = {L dL \over a^2} - {L^2 \over a^3} da + {dV \over dr}(a) da~~\ .
\label{varELa}
\end{equation}
In view of Eq.~(\ref{gravpot}) this may be rewritten
\begin{equation}
da = {a \over v_{\rm rot}^2 - v_\phi(a)^2}
(dE - {L dL \over a^2})
\label{varELa2}
\end{equation}
where $v_\phi(r) = L/r$ is the velocity in the direction of galactic 
rotation of the particles that are in the galactic plane.  The spread
$\delta a$ in caustic ring radius is therefore given by 
\begin{equation}
(\delta a)^2 \equiv <(da)^2> = 
{a^2 \over (v_{\rm rot}^2 - v_\phi(a)^2)^2}
[(\delta E)^2 + {L^2 \over a^4} (\delta L)^2 
- 2 {L \over a^2} <dE~dL>]
\label{crs}
\end{equation}
where, as in Section III, brackets indicate averaging over the 
${d^2N \over dE dL}$ distribution of the dark matter particles,  
$\delta E \equiv \sqrt{<(dE)^2>}$ and $\delta L \equiv \sqrt{<(dL)^2>}$.  
The second term in the square brackets on the RHS of Eq.~(\ref{crs}) 
dominates over the first term since
\begin{equation}
{a^2 \delta E \over L \delta L} = 
{a v_{\rm av} \over v_\phi(a) D_{\rm av}}
\end{equation}
and $v_{\rm av}$ is of order $v_\phi(a)$ whereas $D_{\rm av}$
is much larger than $a$, as was discussed in Section III.  The 
second term also dominates over the third term since
\begin{equation}
<dE~dL>~ \sim ~\delta E ~ \delta L ~{a \over R} ~<< ~\delta E~\delta L~~\ .
\end{equation}
Hence 
\begin{equation}
\delta a \simeq \sigma_v 
{v_\phi(a) D_{\rm av} \over v_\phi^2(a) - v_{\rm rot}^2} 
\simeq {\sigma_v \over 426~{\rm km/s}} D_{\rm av}
\label{almost}
\end{equation}
where we used $v_\phi(a) = 520$ km/s \cite{MWhalo}.  It remains 
to estimate $D_{\rm av}$, the average distance from the galactic
center where the processes took place by which the flow presently 
constituting the 5th caustic ring acquired velocity dispersion.
Of course it is hard to give a precise value.  The present turnaround 
radius of the flow constituting the 5th caustic ring is $R_5$ = 121 kpc 
\cite{MWhalo}. As a rough estimate, we set $D_{\rm av} \sim R_5/2$.  
With $\delta a \lesssim$ 10 pc, this yields
\begin{equation}
\sigma_v \lesssim 71~{\rm m/s}~~\ .
\label{ldv}
\end{equation}
In ref.~\cite{MWcr}, an  upper limit on $\sigma_v$ was estimated 
using  $\delta a \sim {R \over v} \sigma_v$ with $R$ the turnaround 
radius and $v$ the velocity of the flow at the caustic.  This yielded 
$\sigma_v \lesssim 53$ m/s.   Although far more work went into justifying 
Eqs.~(\ref{almost}) and (\ref{ldv}), the two estimates are qualitatively 
consistent since $D_{\rm av} \sim R$ and $v = v_\phi(a)$.  The difference 
between the two estimates may be taken as a measure of the uncertainty on 
the bound.  At any rate, the bound on $\sigma_v$ from the sharpness of 
caustic rings is extraordinarily severe in view of the commonly made 
assertion that the dark matter falling onto a galaxy is in clumps 
with velocity dispersion of order 10 km/s.

To obtain an upper limit on the energy dispersion 
$~\delta E = m v_{\rm av} \sigma_v$ of the flow forming
the 5th caustic ring we set $v_{\rm av} \sim$ 300 km/s.  
This yields
\begin{equation}
{\delta E \over m} \lesssim 2.4 \cdot 10^{-10}~~\ .
\label{bendis}
\end{equation}
If the axion frequency is 1 GHz, as in the example given in 
the Introduction, the upper limit on the widths of the peaks 
associated with the Big Flow and the Little Flow in the cavity 
detector of dark matter axions is of order 0.24 Hz.  Let us 
emphasize that there is nothing to suggest that the upper 
bound is saturated.  

\subsection{Velocity dispersion ellipse of the Big Flow}

In the Caustic Ring Model of the Milky Way halo, the Big Flow 
has a large density on Earth as a result of our proximity to a
cusp in the 5th caustic ring of dark matter.  The inner radius 
of the 5th caustic ring, derived from a rise in the Milky Way 
rotation curve and from the triangular feature in the IRAS map 
of the Galactic plane, is $a$ = 8.31 kpc whereas its outer 
radius is $a + p$ = 8.44 kpc \cite{MWcr}.  See Fig.~2.  These 
values assume that our own distance to the Galactic center, which 
sets the scale, is 8.5 kpc.  Note that the values of the inner 
and outer radii are at the tangent point of our line of sight 
to the 5th caustic ring.  If axial symmetry is assumed, as in the 
Caustic Ring Model, we are just outside the tricusp cross-section
of the 5th caustic ring, near the cusp at the outer radius, which 
we call henceforth the `outer cusp'.  In that case there are two 
flows on Earth associated with the 5th caustic ring, the Big Flow 
and the Little Flow.  The uncertainty on the density of the Big Flow 
is large since it is sensitive to our distance to the outer cusp.  
Axial symmetry need not be a good approximation in this context.
It could be sufficiently broken that we are located inside the 
tricusp instead of just outside.  If we are located inside the 
tricusp, there are four flows on Earth associated with the 5th 
caustic ring.  If we are inside the tricusp and near the outer 
cusp three of the flows are Big Flows and the fourth is the Little 
Flow.  The Little Flow is the same as before.  It does not participate 
in the outer cusp singularity.  The Big Flows do participate in the 
outer cusp singularity. The densities, velocities and velocity 
dispersions of the Big Flows are given by the equations in Section IV 
as a function of position relative to the cusp.  The Big Flows all 
have comparatively large dispersions in the component of velocity 
perpendicular to the symmetry axis of the cusp, i.e. their velocity 
ellipses are elongated in the direction perpendicular to the Galactic 
plane.

\section{Summary}

Motivated by the prospect that dark matter may some day be detected
on Earth, we set out to predict properties of the velocity dispersion 
ellipsoid of the Big Flow.  The Big Flow dominates the local dark matter 
distribution in the Caustic Ring Model of the Milky Way halo due to our 
proximity to the 5th caustic ring of dark matter in our galaxy.  We 
analyzed a cold collisionless stationary flow in one dimension and 
derived how the velocity dispersion changes along such a flow.   In 
one dimension the problem is simple because energy conservation and 
Liouville's theorem control the outcome.  To make headway in two 
dimensions we assumed that the potential in which the particles fall 
is axially symmetric as well as time-independent so that both energy 
and angular momentum are conserved.  We derive the evolution of the 
velocity dispersion ellipse along a stationary axially symmetric flow 
under those assumptions.  The local velocity dispersion always becomes 
large when approaching a caustic because the density becomes large but 
the phase space density is constant.  We introduced a technique for 
obtaining the leading behavior of the velocity dispersion near caustics, 
and applied the technique to fold and cusp caustics.  Finally we used 
our results to obtain an upper limit on the energy dispersion of the 
Big Flow from the observed sharpness of the 5th caustic ring and a 
prediction for the dispersion in its velocity components.

\begin{acknowledgments}

This work was supported in part by the U.S. Department of Energy under 
grant DE-FG02-97ER41209 at the University of Florida, and the National 
Science Foundation under Grant No. PHYS-1066293 at the Aspen Center for 
Physics.  Fermilab is operated by Fermi Research Alliance, LLC, under 
Contract No. DE-AC02-07CH11359 with the US Department of Energy. NB is 
supported by the Fermilab Graduate Student Research Program in Theoretical 
Physics.

\end{acknowledgments}


\newpage



\newpage

\vspace{-3in}
\begin{figure}
\begin{center}
\includegraphics[angle=360, height=100mm]{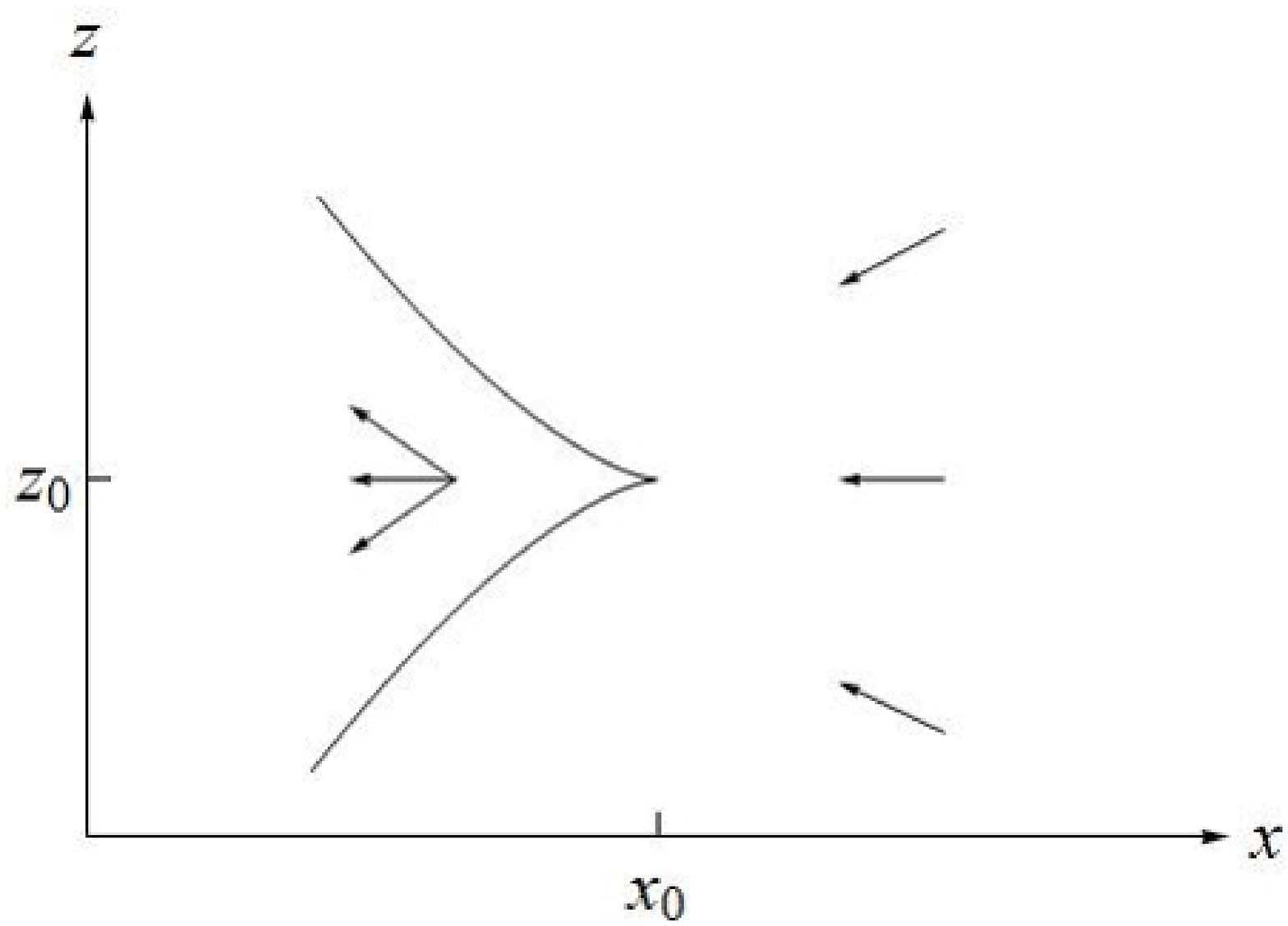}
\vspace{0.3in}
\caption{The curve with the cusp is the location of the fold 
caustic in the flow described by Eqs.~(\ref{cmap}).  The arrows 
indicate local velocity vectors.  There is one flow at every point 
to the right of the curve, and three flows at every point to its
left.}
\end{center}
\label{cusp}
\end{figure} 

\newpage

\vspace{-3in}
\begin{figure}
\begin{center}
\includegraphics[angle=360, height=100mm]{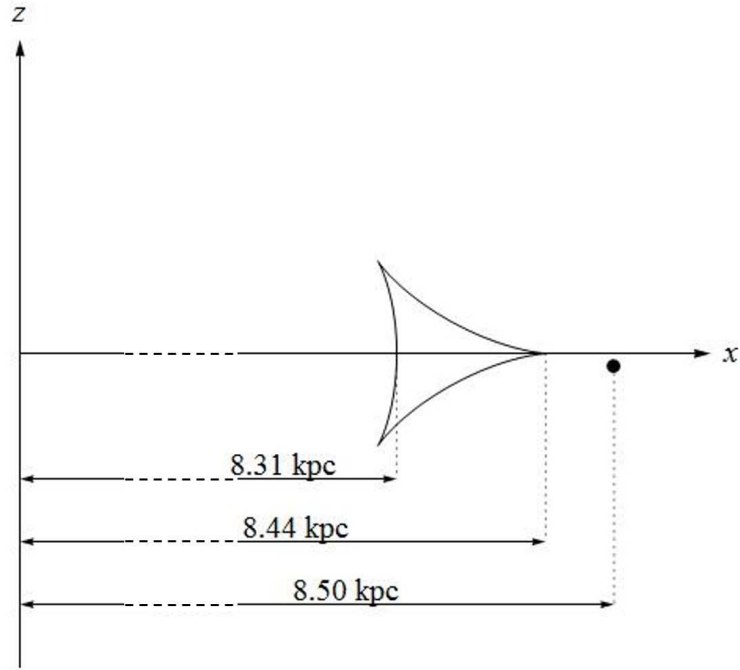}
\vspace{0.3in}
\caption{Relative position of the Sun and the 5th caustic ring
in the Caustic Ring Model of the Milky Way halo.  The Sun's position 
is indicated by the dot.  The $x$-axis is parallel to the Galactic 
plane.  The tricusp shape is the cross-section of the 5th caustic 
ring.  The model assumes axial symmetry.  Because of axial symmetry 
breaking, the size of the tricusp and the position of the Sun relative 
to it may differ from what the figure shows.} 
\end{center} 
\label{tricusp}  
\end{figure}  


\end{document}